\documentclass[preprint,12pt]{elsarticle}

\usepackage{graphicx}

\usepackage{amssymb}

\journal{Solid State Communications}

\begin{document}

\begin{frontmatter}



\title{Pressure cycle of superconducting Cs$_{0.8}$Fe$_{2}$Se$_2$: a transport study}


\author[DPMC]{G. Seyfarth}
\author[DPMC]{D. Jaccard}
\author[CNEA]{P.~Pedrazzini}
\author[PSI1]{A.~Krzton-Maziopa}
\author[PSI1]{E.~Pomjakushina}
\author[PSI1]{K.~Conder}
\author[PSI2]{Z.~Shermadini}

\address[DPMC]{DPMC - Universit\'e de Gen\`eve, Quai Ernest Ansermet 24, 1211 Geneva 4, Switzerland}
\address[CNEA]{Lab.~Bajas Temperaturas and Instituto Balseiro, Centro At\'omico Bariloche (CNEA), 8400 S.C.~de Bariloche, Argentina}
\address[PSI1]{Lab. for Developments and Methods, Paul Scherrer Institute, 5232 Villigen PSI, Switzerland}
\address[PSI2]{Lab. for Muon-Spin Spectroscopy, Paul Scherrer Institute, 5232 Villigen PSI, Switzerland}

\begin{abstract}
We report measurements of the temperature and pressure dependence of the electrical resistivity of single crystalline iron-based chalcogenide Cs$_{0.8}$Fe$_{2}$Se$_2$. In this material superconductivity with $T_{\rm c}\sim 30\,$K develops from a normal state with extremely large resistivity. At ambient pressure a large ``hump'' in the resistivity is observed around $200\,$K. Under pressure, the resistivity decreases by two orders of magnitude, concomitant with a sudden $T_c$ suppression around $p_c\sim 8\,$GPa. Even at $9\,$GPa a metallic resistivity state is not recovered, and the $\rho(T)$ ``hump'' is still detected. A comparison of the data measured upon increasing and decreasing the external pressure leads us to suggest that superconductivity is not related to this hump.
\end{abstract}


\end{frontmatter}


\section{Introduction}\label{sec:intro}

From an structural point of view, FeSe is the simplest material among the new Fe-based pnictide and chalcogenide ``high temperature'' superconductors \cite{Hsu2008,Mizuguchi2010review}. It is a tetragonal compound in which layers formed by edge-sharing FeSe$_4$ tetrahedra are stacked along the $c-$axes of the crystal. It possesses several outstanding characteristics, among them a large pressure effect on the superconducting temperature that increases from $T_{\rm c}(p=0)\approx 13\,$K to $T_{\rm c}\approx 37\,$K at $4\,{\rm GPa}$ \cite{Mizuguchi2008}. This steep raise in  $T_{\rm c}(p)$ is thought to be due in part to a reduction in the distance $h$ between the Se anion and the Fe plane (known as the ``anion height''), that approaches the optimum height for superconductivity $h^*\approx 1.38\,{\rm \AA}$ \cite{Mizuguchi2010}.

Recently, a new family of Fe-based superconductors with $T_{\rm c}\sim 30\,$K and the general formula A$_x$Fe$_{2-y}$Se$_2$ has been identified. Element A is either the alkaline K \cite{Guo2010}, Rb \cite{Wang2010} or Cs \cite{Krzton2011}, or also Tl in the +1 valence \cite{Fang2010}. These compounds crystalize in the well known ThCr$_2$Si$_2$-type tetragonal structure(space group $I4/mmm$), obtained by the intercalation of A in superconducting FeSe. Introducing the element A expands the tetragonal $c-$axis but reduces the anion height, that approaches the optimum value $h^*$ \cite{Krzton2011}. Muon-spin spectroscopy, resistivity, magnetization and differential scanning calorimetry investigations performed on the system A\,=\,Cs (hereafter Cs-122) have shown a microscopic coexistence between superconductivity and a magnetic phase with very high $T_N=478\,$K \cite{Shermadini2011}. A similar behavior has been subsequently reported in the system A\,=\,K, based on a neutron work \cite{Bao2011}. Besides, resistivity measurements on K$_{0.8}$Fe$_{1.7}$Se$_2$ up to $11\,$GPa \cite{Guo2011} seem to indicate that $T_{\rm c}$ cannot be further optimized by applying high pressures and suggest a relationship between the resistivity hump around $200\,$K and the occurrence of superconductivity. Our present results question this relation, in line with recent reports studying this interdependence in a large variety of samples at ambient pressure \cite{Luo2011,Wang2011,Hu2011}.

\section{Experimental Details}\label{sec:experiment}

Single crystals of nominal composition Cs$_{0.8}$(FeSe$_{0.98}$)$_{2}$ were grown using Bridgman technique \cite{Krzton2011}. Detailed crystallographic analysis revealed the presence of only one single phase \cite{Pomjakushin2011}, and magnetization data \cite{Shermadini2011} are compatible with 100\% superconducting volume fraction (on crystals of same batch as the present one). Two samples were carefully cleaved from a larger crystal. High-pressure four-probe resistivity measurements along the basal (ab) plane were performed on sample $S1$, a $1020\times 170\times 30\,\mu{\rm m}^3$ cuboid. The sample together with a strip of Pb, which served as pressure gauge, was mounted in a Bridgman-type high-pressure cell \cite{Jaccard1998} using steatite as transmitting medium. Electrical contacts were obtained by pressing annealed $10\,\mu{\rm m}$ gold wires directly
onto the sample. We have observed a progressive reduction of the wire-sample contact resistance, from
around $1\,{\rm k}\Omega$ at $1.2\,$GPa to $\sim 10\,\Omega$ at $9\,$GPa. All measurements have been performed
with currents ranging from $1\,\mu{\rm A}$ to $10\,\mu{\rm A}$ ($0.02\le j \le 0.2\,{\rm A}\,{cm}^{-2}$),
although larger currents do not have a measurable effect on the superconducting transition (the reported critical current density in K$_{0.86}$Fe$_{1.84}$Se$_2$ is of the order of $10^3\,{\rm A}\,{cm}^{-2}$ \cite{Hu2011}).

In order to test the relation between superconductivity and the normal-state transport of Cs-122, we have studied the temperature dependence of the electrical resistivity with both increasing and decreasing pressure.

A $2.0\times 0.72\times 0.36\,{\rm mm}^3$ cuboid, sample $S2$, was measured at zero pressure. Electrical contacts were obtained by gluing gold wires with the DuPont conductor paste 4929N.

\section{Results}\label{sec:results}

\begin{figure}[htb]
\begin{center}
\includegraphics[width=0.55\textwidth]{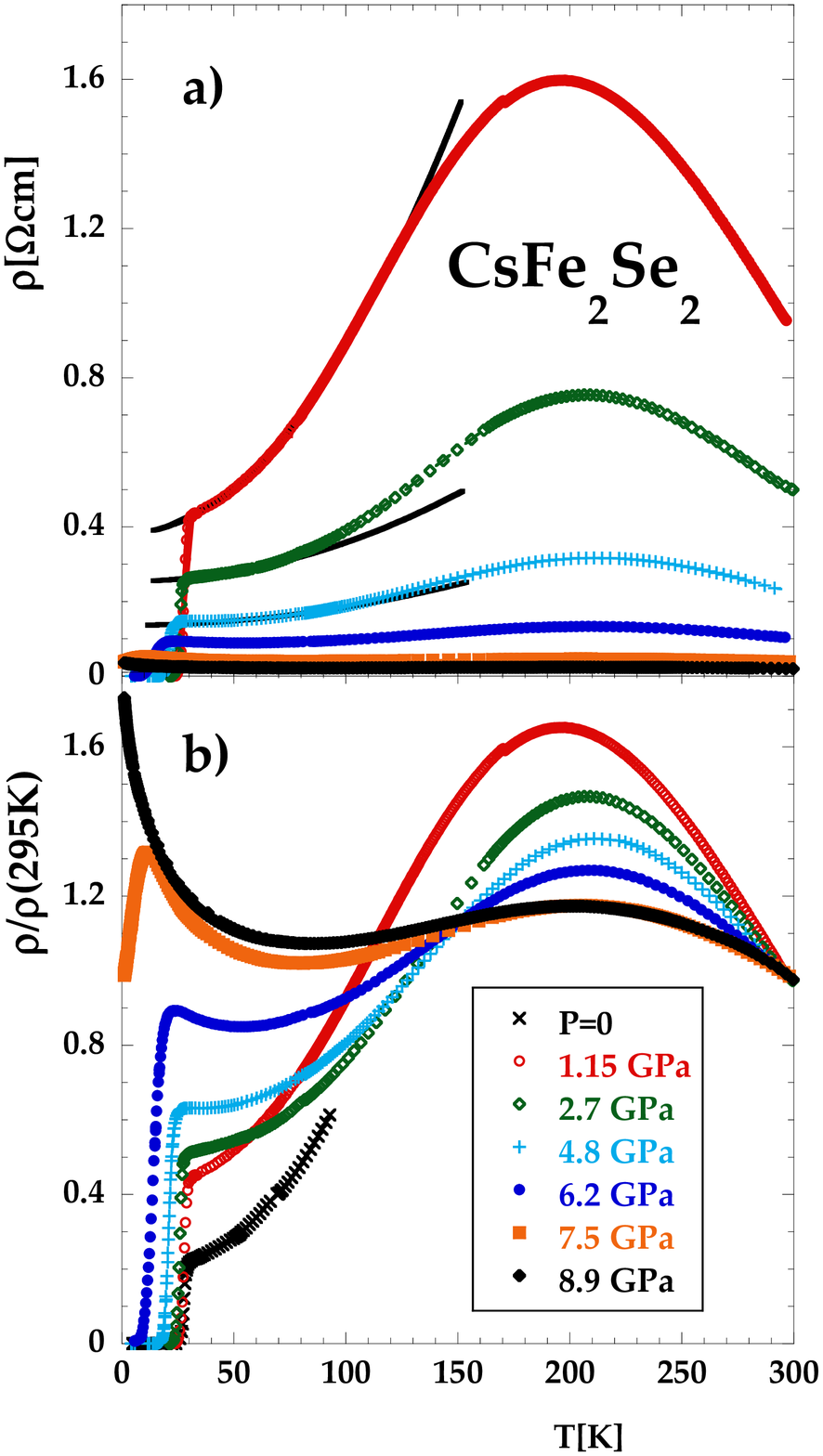}
\caption{Temperature dependence of the electrical resistivity of Cs-122 at different pressures (increasing). a) Raw data (with $T^2$-fits as continuous lines). b) Data normalized to the corresponding room temperature value. Suppression of $T_c$ and of the high resistivity with increasing pressure can clearly be observed. \label{fig:rhovst-pup}}
\end{center}
\end{figure}

In Fig.~\ref{fig:rhovst-pup}-a we present the temperature dependence of the electrical resistivity of sample $S1$ at selected {\em increasing} pressures ($p\uparrow$). At the lowest pressure, $p\approx 1.2\,$GPa, $\rho(T)$ displays a similar $T-$dependence as sample $S2$ at $p=0$, which is typical of other A$_x$Fe$_2$Se$_2$ compounds reported in the literature \cite{Guo2010,Hu2011,Lei2011,Wang2010,Shermadini2011,Wang2011thermo}. The resistivity first increases towards lower temperatures, it displays a ``hump'' at $T_{\rm max}\sim 200\,$K and the superconducting transition at $T_{\rm c}^{50\%}\approx 28\,$K. Between the onset of SC, $T_{\rm c}^{\rm onset}\approx 30\,$K, and roughly $90\,$K the resistivity can be described as $\rho(T)=\rho_0+\alpha T^2$, with $\rho_0=0.38\,\Omega\,{\rm cm}$ and $\alpha=51\,\mu\Omega\,{\rm cm}/{\rm K}^2$ (at 1.2\,GPa), see the fits in Fig.~\ref{fig:rhovst-pup}-a. Between $p=0$ and $4.8\,$GPa the resistivity coefficient decreases roughly as $\alpha(p)\sim 120 e^{-p/1.2}\,\mu\Omega\,{\rm cm}/{\rm K}^2$. For $p>5\,$GPa this $T^2$ behavior is observed in a very narrow temperature range.

As pressure increases we observe a dramatic drop in the magnitude of the electrical resistivity. Despite the large reduction of $\rho(T)$, no metallic behavior is observed even at $p\approx 9\,$GPa and the hump remains at roughly the same temperature, $T_{\rm max}\approx 200\,$K, although with a smaller relative amplitude. This can be seen in Fig.~\ref{fig:rhovst-pup}-b depicting the resistivity normalized at $295\,$K, $\rho(T)/\rho(295\,{\rm K})$. The ratio $\rho(T_{\rm max})/\rho(295\,{\rm K})$ decreases from $1.65$ at $1.15\,$GPa down to $1.17$ at $8.9\,$GPa. If $T_{\rm max}$ is associated with some characteristic temperature of this material, its roughly constant value indicates that it is related to a process weakly affected by pressure.

\begin{figure}[htb]
\begin{center}
\includegraphics[width=0.55\textwidth]{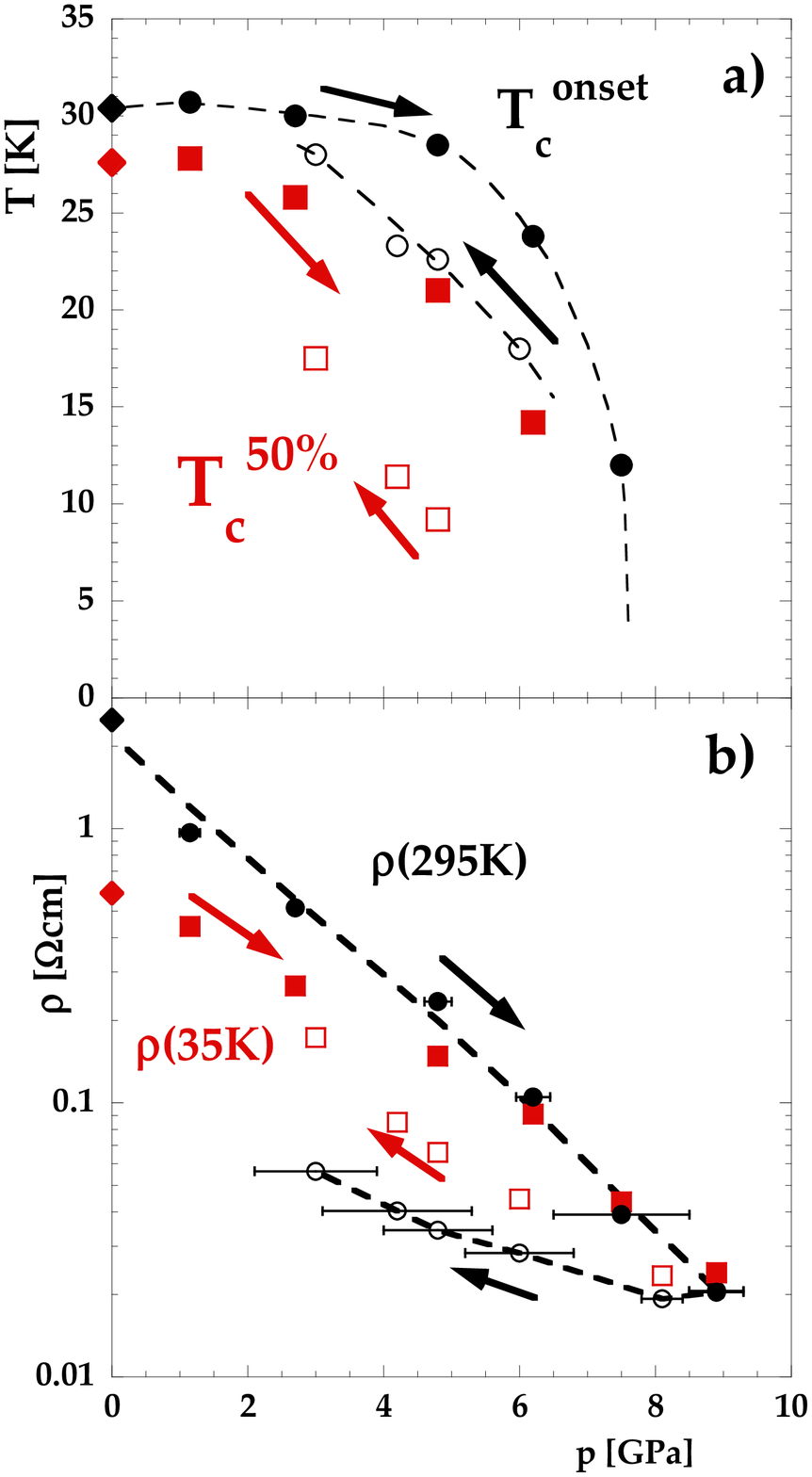}
\caption{Pressure dependence of (a) the superconducting transition temperature, as seen by resistivity (black: onset and red: mid-height, 50\%), and of (b) the resistivity at 35 (red) and 295K (black). Full symbols indicate increasing, and open symbols decreasing pressure. The indicated error bars mainly reflect the width of the superconducting transition of Pb. Obviously, the onset $T_c$ is almost reversible (with hysteresis), whereas resistivity does not recover the initially very high magnitude. \label{fig:PhaseDiag}}
\end{center}
\end{figure}

The progressive decrease of $T_{\rm c}^{\rm onset}(p)$ can be followed in Fig.~\ref{fig:rhovst-pup}-b. For comparison, we have included the low temperature $\rho(T)$ data measured at $p=0$ in sample $S2$. We see that $T_{\rm c}(p)$ remains almost unchanged between $p=0$ and $\sim 3\,$GPa and that it decreases steadily for $p>4\,$GPa. At $p\sim 7.5\,$GPa we only detect a partial transition while at $8.9\,$GPa $\rho(T)$ increases continuously as $T\to 1.2\,$K. The evolution of $T_{\rm c}^{\rm onset}(p)$ and $T_{\rm c}^{50\%}(p)$ is depicted in the phase diagram of Fig.~\ref{fig:PhaseDiag}-a. We have not detected any anomaly in $\rho(T)$ that may indicate that the magnetic order measured in Cs-122 at $T_{\rm N}\approx 480\,$K \cite{Shermadini2011,Pomjakushin2011} is being suppressed in this pressure range.

\begin{figure}[htb]
\begin{center}
\includegraphics[width=0.75\textwidth]{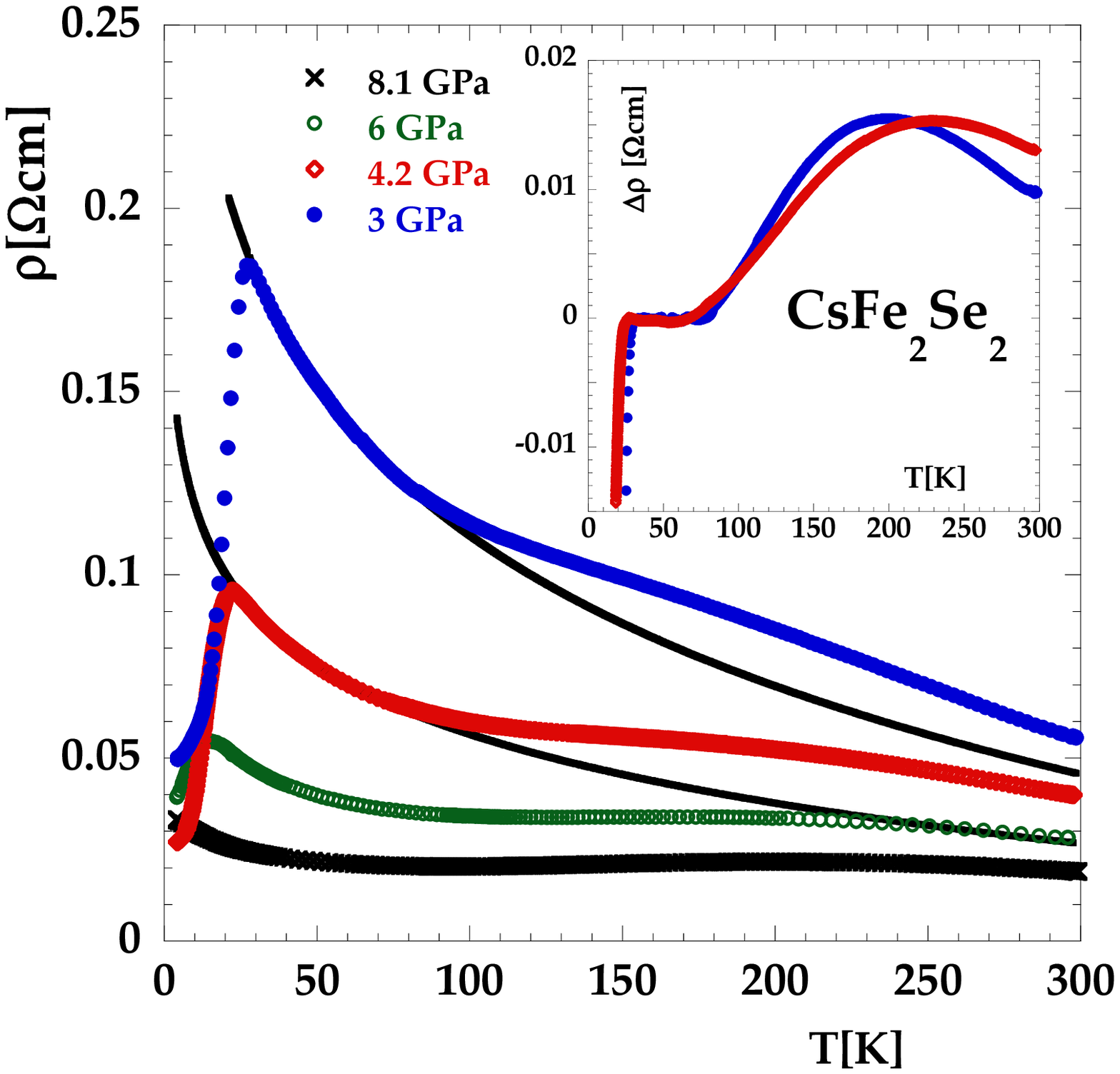}
\caption{Temperature dependence of the electrical resistivity of Cs-122 at different pressures (decreasing). The onset of $T_c$ recovers, while the hump is much less reversible. Solid lines are a fit of the type $\alpha \ln(T_0/T)$, where $\alpha$ and $T_0$ are the adjusted parameters. The inset depicts the difference of the measured resistivity with these fits, $\rho(T)-\alpha \ln(T_0/T)$ , in order to make the small, still existing hump visible. \label{fig:rhovst-pdown}}
\end{center}
\end{figure}

Figure \ref{fig:rhovst-pdown} depicts $\rho(T)$ of Cs-122 measured with {\em decreasing} pressure ($p\downarrow$). The $\rho(T)$ dependence is notably changed when compared with the data presented in figure \ref{fig:rhovst-pup}. The hump is no longer evident for $p\downarrow<6\,$GPa and $\rho(T)$ increases continuously towards lower temperatures. For $p\downarrow = 3\,$GPa and $4.2\,$GPa the resistivity follows $\rho(T)\sim \ln(T_0/T)$ between $T_{\rm c}^{\rm onset}$ and roughly $80\,$K, see the fits represented by continuous curves. The inset displays $\rho(T)-\alpha \ln(T_0/T)$ for these two pressures. The maximum is still detected around $200\,$K, but with a very small amplitude. Despite this we detect a partial superconducting transition with an onset that reaches $T_{\rm c}^{\rm onset}\approx 30\,$K at $3\,$GPa, coinciding with the onset temperature observed for $p\uparrow\approx 2.7\,$GPa: the superconducting state recovers, while the hump attains only $\sim 5\%$ of its previous value at a similar pressure. At least phenomenologically, this superconducting state develops in a material with semiconductor-like behavior.

In Fig.~\ref{fig:PhaseDiag}-a we present the $p-T$ phase diagram of Cs-122, comparing the changes in the superconducting ordering temperature with increasing and decreasing pressure. The circular symbols represent the onset of superconductivity, $T_{\rm c}^{\rm onset}(p)$, while the rectangular symbols indicate the mid-point of the resistive transition, $T_{\rm c}^{50\%}(p)$. We first focus on the ``{\em increasing} pressure'' experiment, represented in the figure by filled symbols. At low pressures, the transition is relatively narrow so $T_{\rm c}^{\rm onset}$ and $T_{\rm c}^{50\%}(p)$ are close and our data agree with the results reported in Ref.~\cite{Ying2011}. With increasing $p$, $T_{\rm c}^{50\%}(p)$ diminishes although $T_{\rm c}^{\rm onset}(p)$ remains almost constant up to $p\sim 5\,$GPa. For $p>5\,$GPa $T_{\rm c}^{\rm onset}(p)$ decreases steeply and it is no longer detected above $8\,$GPa. We estimate a critical pressure for superconductivity $p^{\rm Cs}_{\rm c}\approx 7.6\,$GPa. The steep decrease of $T_{\rm c}^{\rm onset}$ as $p\to p^{\rm Cs}_{\rm c}$ contrasts with the progressive suppression of superconductivity in isostructural K$_{0.8}$Fe$_{1.7}$Se$_{2}$ \cite{Guo2011}, although the critical pressure $p^{\rm K}_{\rm c}\approx 9\,$GPa is rather similar. Notice
that this pressure should correspond to a large volume reduction of the order of $15\%$ \cite{volume}.

With {\em decreasing} pressure (open symbols) $T_{\rm c}^{\rm onset}$ recovers completely around $3\,$GPa, although the transition is rather broad and no zero-resistance state is detected. We interpret this recovery as a reversibility of superconductivity under pressure cycling despite the apparent pressure hysteresis of roughly $2\,$GPa. In Fig.~\ref{fig:PhaseDiag}-b we represent the $\rho(35\,{\rm K})$ and $\rho(295\,{\rm K})$ $p$-dependence to stress that the resistivity is not reversible at all.

\section{Discussion}\label{sec:discussion}

As already pointed out, superconductivity develops in AFe$_2$Se$_2$ compounds out of normal state resistivities of $\sim0.2\,\Omega\,{\rm cm}$, that is about 3 orders of magnitude higher than say in FeSe \cite{Hsu2008}. For the Cs-122 samples studied here, we estimate $\rho(295\,{\rm K})\approx 2.5\,\Omega\,{\rm cm}$ at $p=0$. We verified that this value does not change while exposing the sample to the air for around one hour, the time necessary to cleave and close the high pressure cell. To our knowledge, there are no other systems that exhibit such a large $T_c$ combined with so large normal state resistivities. Even more, Hall effect data in a related compound do not point to a reduced carriers number \cite{Guo2010}. On the other hand, it should be noted that the AFe$_2$Se$_2$ family is in close proximity to an insulating state \cite{Fang2010,Wang2011,Luo2011}. Further, it is remarkable that the room temperature resistivity decreases two orders of magnitude between $p=0$ and $9\,$GPa, see figure \ref{fig:PhaseDiag}-b (semi-log scale). Nevertheless, even at the highest pressure ($9\,$GPa) we do not observe a metallic behavior of the resistivity.

From our previous studies of organic superconductors on different pressure media \cite{Ruetschi2007} there is strong indication that the limited pressure conditions in the steatite medium may induce the resistivity upturn observed at low temperature. The fact that this feature steadily increases in magnitude with the number of pressure steps supports this point of view. So it remains a challenging question to determine whether the high pressure phase is intrinsically non-metallic. High pressure experiments under better hydrostatic conditions are highly desirable.

\begin{figure}[htb]
\begin{center}
\includegraphics[width=0.55\textwidth]{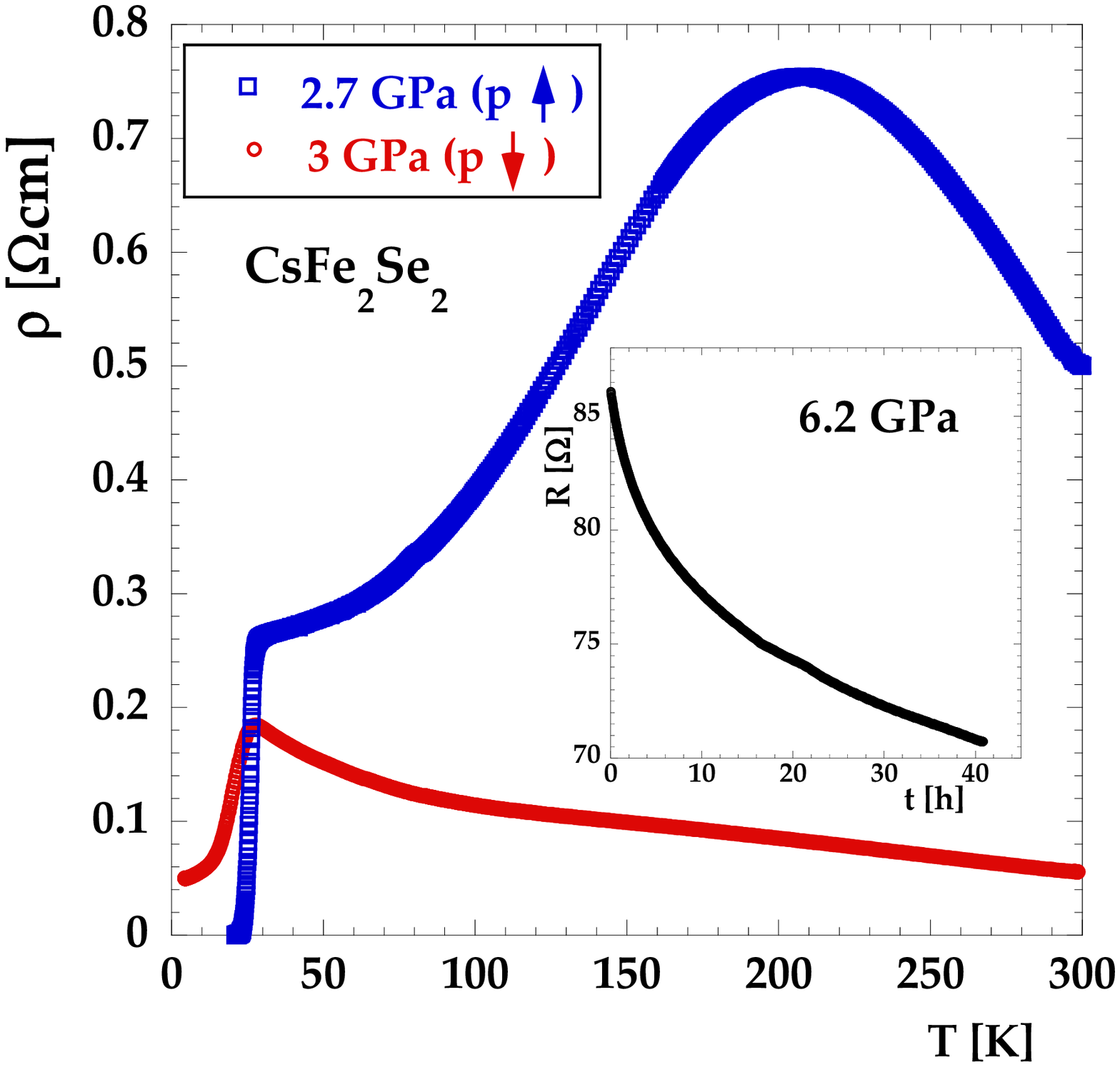}
\caption{Temperature dependence of the electrical resistivity of Cs-122 at around 3\,GPa, measured when increasing and decreasing pressure. The onset of $T_c$ seems quite robust against pressure cycling, whereas the complete $\rho (T)$ curve is strongly modified. The inset visualizes the time dependence (at room temperature) of the electric resistance of the sample after increasing the pressure from 4.8 to 6.2\, GPa: the observed changes are not negligible. \label{fig:compare}}
\end{center}
\end{figure}

Concerning the resistivity hump, let us first recall that experimentally it is strongly affected by pressure-cycling: Figure \ref{fig:compare} compares resistivity data at around 3\,GPa for increasing and decreasing pressure. On the bottom curve ($p\downarrow$) one can only see a trace around 200\,K of the resistivity hump and the development of the low temperature upturn (see also Fig.~\ref{fig:rhovst-pdown}), while the onset temperature of superconductivity recovers and is close to that when increasing pressure ($p\uparrow$, upper curve). As an explanation, one possible scenario evokes the ordering of iron vacancies in the FeSe layers just above $T_N$ \cite{Wang2011,Pomjakushin2011,Fang2010,Liu2011,Bao2011,Wang2011super,Zavalij2011,Bacsa2011}, which could be pressure-sensitive and not reversible. The inset of Fig. \ref{fig:compare} exhibits the time evolution of the sample resistance at room temperature, just after the increase of pressure to $6.2\,$GPa. A significant relaxation of $\sim15\%$ in 24 hours is observed. We have confirmed that $\rho(T)$ is $T-$reversible, meaning that this mechanism is not active at low temperatures. The observed relaxation could be related to a mechanism of moving vacancies. Whatever the microscopic origin of the resistivity hump, phenomenologically it is now clearly established from detailed composition studies that transport properties strongly depend on the Fe (and A) content in the non-stoichiometric A$_x$Fe$_{2-y}$Se$_2$ compounds \cite{Fang2010,Wang2011,Hu2011,Luo2011,Bao2011_2}: in particular, decreasing the Fe deficiency shifts the hump to higher temperatures, whereas the superconducting $T_c$ is hardly affected. Combined with our results obtained after one pressure cycle (full recovery of $T_c$ but only partial of the hump), a close relationship between the resistivity hump and superconductivity seems unlikely, in contrast to the results reported in \cite{Guo2011}.

Finally, doping studies in AFe2Se2 compounds show that $T_c$ appears rather abruptly around $30\,$K \cite{Fang2010}. The steep decrease of $T_{\rm c}^{onset}$ in a narrow pressure range just before $p_c$ as well as the $T_c(p)$ hysteresis suggest that the $T_c$ vanishing might not be a continuous phase transition. In any case, the apparent ``robustness'' of ``30~K-superconductivity'' in the AFe$_2$Se$_2$ family should be noted: roughly speaking, superconductivity seems to appear irrespectively of A, doping\footnote{except for Co-doping \cite{Zhou2011}, quenching $T_c$ abruptly (quite opposite to the case of BaFe$_2$As$_2$)} and stoichiometry (within the appropriate parameter range) with a somehow ``fixed'' transition temperature. Even pressure-cycling seems inefficient to significantly affect $T_c$ (except to make it disappear promptly above 4~GPa for Cs-122).

\section{Conclusion}\label{sec:conclu}

In conclusion, high pressure measurements on single crystalline Cs$_{0.8}$Fe$_{2}$Se$_2$ show a suppression of $T_c(p)$, which is almost constant up to $5\,$GPa and then it decreases steeply, not being detected above $8\,$GPa. The resistivity hump of unknown origin is only very partially recovered with decreasing pressure, while $T_c$ seems to be reversible. This questions the connection between $T_c$ and the resistivity hump, in line with recent publications \cite{Luo2011,Wang2011,Hu2011} that arrive at a similar conclusion. At the high pressure side of the SC-phase a metallic state is not unambiguously recovered. To test this, experiments in a more hydrostatic pressure medium should be performed.

\section{Acknowledgements}

This work was supported by the Swiss National Science Foundation through the NCCR ``MaNEP''. P.P. is a member of CONICET.






\begin{thebibliography}{99}


\bibitem{Hsu2008} F.-C.~Hsu {\em et al.}, Proc.~Nat.~Acad.~Sci.~U.S.A.~{\bf 105} (2008) 14262.
\bibitem{Mizuguchi2010review} Y.~Mizuguchi {\em et al.}, JPSJ~{\bf 79} (2010) 102001. (review article)
\bibitem{Mizuguchi2008} Y.~Mizuguchi {\em et al.}, Appl.~Phys.~Lett.~{\bf 93} (2008) 152505.
\bibitem{Mizuguchi2010} Y.~Mizuguchi {\em et al.}, Supercond.~Sci.~Technol.~{\bf 23} (2008) 054013.
\bibitem{Guo2010} J.~Guo {\em et al.}, Phys.~Rev.~B {\bf 82} (2010) 180520.
\bibitem{Wang2010} A.F.~Wang {\em et al.}, arXiv cond-mat 1012.5525.
\bibitem{Krzton2011} A.~Krzton-Maziopa {\em et al.}, J.~Phys.: Condens.~Matter {\bf 23} (2011) 052203.
\bibitem{Fang2010} M.~Fang {\em et al.}, arXiv cond-mat 1012.5236.
\bibitem{Shermadini2011} Z.~Shermadini {\em et al.}, arXiv cond-mat 1101.1873.
\bibitem{Bao2011} W.~Bao {\em et al.}, arXiv cond-mat 1102.0830.
\bibitem{Guo2011} J.~Guo {\em et al.}, arXiv cond-mat 1101.0092.
\bibitem{Luo2011} X.G.~Luo {\em et al.}, arXiv cond-mat 1101.5670.
\bibitem{Wang2011} D.M.~Wang {\em et al.}, arXiv cond-mat 1101.0789.
\bibitem{Hu2011} R.~Hu {\em et al.}, arXiv cond-mat 1102.1931.
\bibitem{Pomjakushin2011} V.Y.~Pomjakushin {\em et al.}, arXiv cond-mat 1102.1919.
\bibitem{Jaccard1998} D.~Jaccard {\em et al.}, Rev.~High Pressure Sci.~Technol.~{\bf 7} (1998) 412.
\bibitem{Lei2011} H.~Lei {\em et al.}, arXiv cond-mat 1102.1010.
\bibitem{Wang2011thermo} K.~Wang {\em et al.}, arXiv cond-mat 1102.2217.
\bibitem{Ying2011} J.J.~Ying {\em et al.}, arXiv cond-mat 1101.1234.
\bibitem{volume} To estimate this value we consider the structural data measured on FeSe by S.~Margadonna {\em et al.}, Phys.~Rev.~B {\bf 80} (2009) 064506.
\bibitem{Ruetschi2007} A.S.~R\"uetschi {\em et al.}, Rev.~Sci.~Instrum.~{\bf 78} (2007) 123901.
\bibitem{Liu2011} R.H.~Liu {\em et al.}, arXiv cond-mat 1102.2783.
\bibitem{Wang2011super} Z.~Wang {\em et al.}, arXiv cond-mat 1101.2059.
\bibitem{Zavalij2011} P.~Zavalij {\em et al.}, arXiv cond-mat 1101.4882.
\bibitem{Bacsa2011} J.~Bacsa {\em et al.}, arXiv cond-mat 1102.0488.
\bibitem{Bao2011_2} W.~Bao {\em et al.}, arXiv cond-mat 1102.3674.
\bibitem{Zhou2011} T.~Zhou {\em et al.}, arXiv cond-mat 1102.3506.

\end{thebibliography}



\end{document}